\documentclass[
aps,%
12pt,%
final,%
notitlepage,%
oneside,%
onecolumn,%
nobibnotes,%
nofootinbib,%
superscriptaddress,%
noshowpacs,%
]%
{revtex4}
\usepackage{graphicx}

\begin{document}

\title{Historical light curve of the black hole binary V4641 Sgr\\
based on the Moscow and Sonneberg plate archives}
\author{\firstname{E.~A.}~\surname{Barsukova}}
\affiliation{
Special Astrophysical Observatory, Russian Academy of Sciences,
Nizhniy Arkhyz, Karachai-Cherkesia, 369167, Russia
} \email{bars@sao.ru}

\author{\firstname{V.~P.}~\surname{Goranskij}}
\affiliation{Sternberg Astronomical Institute, Moscow University,
Moscow, 119992 Russia}

\author{\firstname{P.}~\surname{Kroll}}
\affiliation{Sternwarte Sonneberg, Sternwartestrasse 32, 
D 96515, Sonneberg, Germany}

\begin{abstract}
\vspace{5mm}
We performed digital processing of a large set of photographic plates
for X-ray binary system V4641 Sgr containing a black hole.
A total of 277 plates were found in archives dated between 1960 and
1992. Photographic observations revealed lower level of outburst
activity if to compare with CCD data after the large 1999 outburst.
Only single outburst which happened in 1978 was confirmed.
If archive photographic data are used along with contemporary
CCD observations, the orbital period may be improved, and its value
is 2.81728 $\pm$0.00004 day.
\end{abstract}

\maketitle

\small{{\bf Keywords:} photometry, X-ray binaries, black holes;
individual: V4641 Sgr}

\section{INTRODUCTION}
V4641 Sgr is a system containing a B9--A0 III star with the mass
of 2.9 $\pm$0.4 M$_{\odot}$ and a black hole of 6.2 $\pm$0.7 M$_{\odot}$
\cite{MDea14}. The orbital period of the system is P = 2.81678 day
\cite{OKKea01}. Its distance is 6.2 $\pm$0.7 kpc \cite{MDea14}.
The star was discovered by Goranskij \cite{Gor78} during the 1978 outburst
but at first it was misidentified with Luyten's variable GM Sgr,
a Mira type star \cite{Gor90}. In early 1999 V4641 Sgr was detected
in X-rays as SAX 1819.3--2525, and later, in 1999 September it
experienced a large outburst in all ranges of electromagnetic
wavelengths and reached magnitude of 8$^m$.9 in $V$ band. In this event,
relativistic jets were detected with the VLA radio interferometer
\cite{HRH00}. The orbital period was improved in \cite{Gor01} by
using photometry. Photometric observations show active states and
outbursts, short-time flaring in the scale of seconds, a temporal
appearance of reflection effect \cite{UKI04,GBB03,GBB07}. Spectroscopy
reveals changes in the profiles of Balmer lines and other
manifestation of the black hole activity.

\section{Archive search}

We found 266 plates of V4641 Sgr in the Moscow plate collection
of the Sternberg Astronomical Institute taken with 40 cm astrograph
of SAI Crimean Station which are dated between 1960 and 1992. Typical
exposures were of 45 minutes. The photographic observations were
ended in 1992 as the production of astronomical plates was stopped.
Partially, the star was estimated by eye on the basis of this set
\cite{Gor90}. Additionally, 11 plates were found in the Sonneberg
collection. Plates were taken between 1984 and 1988 with the similar
astrograph. The exposure time of these plates was 30 minutes.
All the plates were produced by AGFA/ORWO factory in Germany.
Plates are sensible to blue light and realize Johnson $B$ band well.
The region of the variable star in these plates including close
outskirts with comparison stars was digitized using a simple convex
lens and FinePix-10 FujiFilm camera in the grey mode. This camera
is self-focused on each frame, and it is sensible enough to take
sharp and clear images with small exposures without any support stand.
The frames in JPEG format given by the camera were transformed to
BITMAP files, and reduced with the WinPG software developed by
V.~Goranskij. We used 17 comparison stars. Characteristic curves were
fitted with the first or second order polynomial. The accuracy of the
fit calculated as mean-squire residual from the polinomial varies
between 0$^m$.03 and 0$^m$.20, mean value is 0$^m$.08. Since 1994,
we continued photoelectric and CCD observations with different
telescopes and devices. CCD observations are performed up to now.
So, our time set of V4641 Sgr continues about half century.

\section{Results}

The total light curve of V4641 Sgr in the $B$ band is presented
in Fig.~1. In the period between 1972 and 1992, only single 1978
outburst was detected, this is the outburst which led to the discovery
of the star. In the peak of the 1978 outburst, the star reached
12$^m$.12 $B$, and the subsequent decay continued about 2 days.
Additionally, a short-time outburst of V4641 Sgr with the 5.87 day
decay in 1901 is found by J.~Grindley and colleagues based on the
Harvard collection, and an episode of a bright state on June 28, 1965
is caught by R.~Hudec and his colleagues using the Bamberg archive.
The last two events were reported at this workshop. One should note
that the object was not detected in X-rays till the 1999 outburst.
After the strong 1999 outburst, the optical outbursts became frequent.
They repeated in 2002, 2003, and 2004. The object appeared in X-rays
also in 2008, 2010, and in 2014.

\begin{figure*}
\includegraphics[scale=0.65]{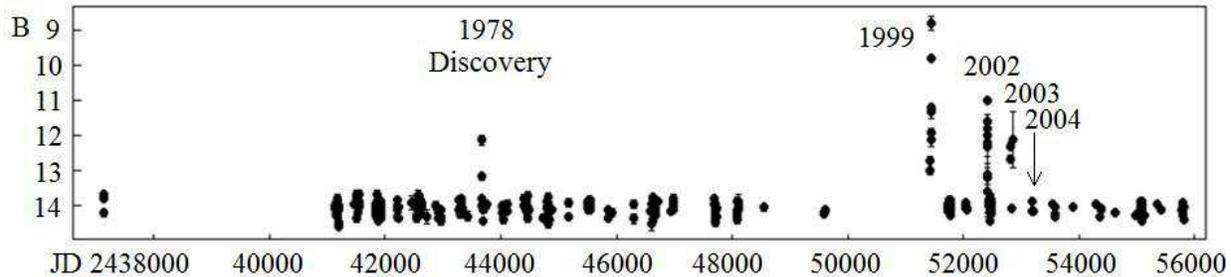}
\caption{Historical half-century light curve of V4641 Sgr.
}
\end{figure*}

Frequency analysis with the phase-dispersion minimization method
\cite{LK65} gives the best period 2$^d$.81728 $\pm0^d.00004$, and
the double-wave light curve with unequal minima depths.
The deepest minimum coincides with the black hole inferior conjunction.
The date of the conjunction is JD Hel. 2451764.337 $\pm$0.005.
The light curve plotted versus phase of this period is given in Fig.~2.
The averaged orbital light curve with the amplitude of 0$^m$.45 in
the primary minimum demonstrates the greatest observed ellipsoidal
effect of stellar component. The depth of the secondary minimum is
0$^m$.28.

\begin{figure*}
\includegraphics[scale=0.85]{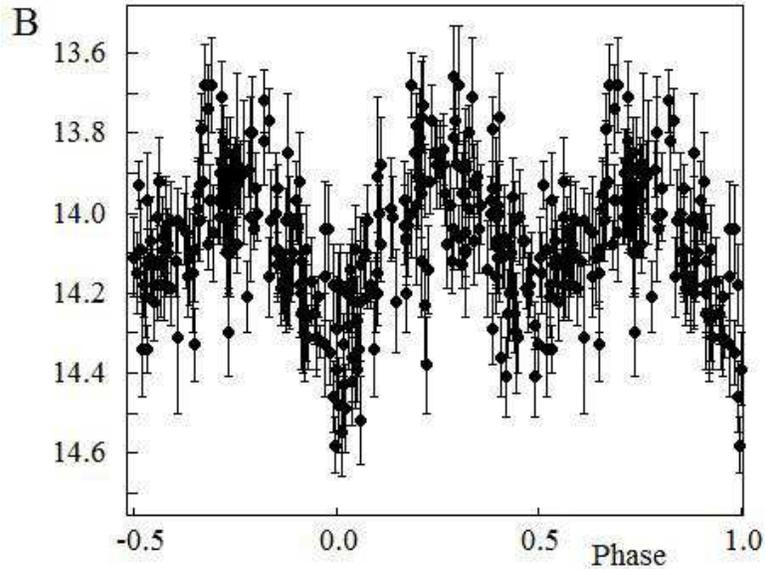}
\caption{Photographic light curve of V4641 Sgr plotted versus
the orbital phase. A primary minimum corresponds with the inferior
conjunction of the black hole.
}
\end{figure*}

We performed CCD $B$, $V$ and $R_C$ monitoring of V4641 Sgr in 16
nights between June 5 and 25, 2007 using SAI Crimean Station 60 cm
telescope and SAO 1 m telescope \cite{GBB07}. These observations
(Fig.~3) show an essential light excess over the quiet light level
visible only in the orbital phases between --0.25 and +0.25 with
the maximum value of 0.15 mag in the $V$ band at the black hole inferior
conjunction. The excess is absent in other orbital phases. This
phenomenon has not been observed previously. We treat it as an
irradiation of the area of A0 type star facing to the black hole.
The area re-emits weak X-ray radiation of the black hole to optical
bands. At the same time with the reflection effect, the radiation at
0.5 -- 10 keV was detected simultaneously with Swift XRT at the
level of (1.2 -- 2.1)$\cdot 10^{-11}$ erg~cm$^{-2}$s$^{-1}$
\cite{CM07}.
\newpage

\begin{figure*}
\includegraphics[scale=0.65]{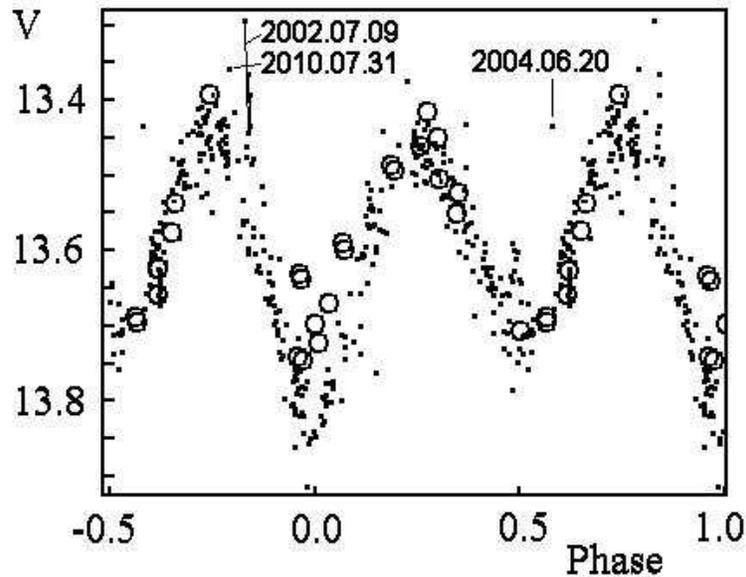}
\caption{
Appearance of the reflection effect in May 2007 (blank circles) in the
$V$ band. Swift detected a weak and variable X-ray radiation. Short
solid lines mark splashes of an optical radiation.
}
\end{figure*}

On July 11, 2001, near the inferior conjunction of the black hole,
we carried out the spectroscopic observations of V4641 Sgr with the
Russian 6 m telescope in order to view the black-hole surroundings
against the stellar disk of the normal component. A depression with
equivalent width EW = 0.5 \AA\ was observed in the red wing of
the H$_{\alpha}$ profile with the maximum absorption at the
heliocentric velocity of 642 km/s \cite{GBB03}. We suggested that
this gas stream could be a part of a rarefied gaseous disk around
the black hole, in the system's orbital plane. Assuming that
the flow is moving through the circular Keplerian orbit around the
black hole, the mass of the black hole is determined as
M$_{BH}$ = 7.1 -- 9.5 M$_{\odot}$, what overlaps the mass range given
by Orosz et al. \cite{OKKea01}.

We explain the deeper minimum in the light curve at the phase of
the black-hole inferior conjunction as an extreme case of
fon Zeipel effect at the surface of the optical companion faced to
the black hole due to very low gravity at this area of the star.
This effect is clearly recorded with our photographic observations.
Photographic observations revealed lower level of outburst activity
compared with CCD data after the large 1999 outburst. Unfortunately,
such low-amplitude effects as X-ray irradiation and rapid flaring
in the scale of seconds are lost due to errors of photographic
observations and long-time exposures.

\end{document}